\documentclass[12pt]{article}
\usepackage[cp1251]{inputenc}
\usepackage{graphicx}
\usepackage{amsfonts,amssymb,eucal,amsmath}

\begin{document}


 \parskip 0pt

\begin{center}
{\bf \Large 
Effects of many-electron atom polarization  in
electron-hole formalism}
\\
{\it  
Halina V. Grushevskaya$^{1}$ and Leonid I. Gurskii$^{2}$
}\\[0.2cm]
 {\it $^{1}$ Physics Department, Belarusian State University,\\
 4 Nezavisimosti Av., 220030 Minsk, BELARUS \\
 $^{2}$ Belarusian State
 University of Informatics  and Radio electronics,\\
 6 P. Brovka Str., 220027 Minsk, BELARUS}\\[0.2cm]
{E-mail: grushevskaja@bsu.by}
\end{center}

\begin{quotation}

{ \bf \centerline{Abstract}  }

The method has been developed to calculate effects of polarization
not only for a atomic core in a field of valent electron, but also
polarization of atom as a whole in the electron-hole formalism.
A secondary
quantized  density matrix for many-electron system was used to find  the
Green function of
a quasiparticle and its effective mass due to many-particle effects.
\end{quotation}

\
\section{Introduction 
}

%
As is known \cite{Fock-foundation}, for odd number of electrons the
Hartree - Fock  method of self-consistent field gives  the
equation, calculation with which help is connected to a problem of
diagonalizing a matrix Lagrange multiplier. It results to possible
noncommutativity of a Fock operator with the operator density matrix
 (a projector on subspace of orbitals for  electron in a
atomic core) and, hence, to absence of an interpretation of the
Fock operator as a Hamiltonian for  single-particle state \cite
{Phillips}, \cite {Weeks}, \cite {Dixon}.
%
%
 The method of
pseudo-potential allows to interpret diagonal elements of the
Lagrange multiplier in a Hartree - Fock equation as the energy of
single-particle state. However, a pseudo-valent orbital constructed
by the method  appears so compressed one, that results to very
underestimated lengths of  chemical bonds in comparison with
experiments  \cite {Durand}, \cite {Christiansen}.
%
%
It, seemingly, is stipulated by the incorrect description of
polarization effects in an atom as in the method of pseudo-potential
one uses approximation of spherically symmetric unexcited
atomic core ("frozen" \ core). Therefore, at the last time in the
\textit {
"ab initio" }   \ calculations fulfilled  the  pseudo-potential
is utilized only within the limits of core, supplementing with the
phenomenological potential describing polarization of the core in
the classical way \cite {Muller}, \cite {Fuentealba}.
%
%
Although, for example, for halogen dimers a relativistic
calculation taking into account the dipole polarizability of
atomic cores
 gives lengths of chemical bounds matching satisfactory  with
experiment data, the  found binding energy  is underestimated with
respect to its real value \cite {Dolg-1}, \cite {Dolg-2}, \cite
{Dolg-3}. Obviously, it is due  to the fact that the correct quantum
description of polarization of an atom is possible, if one assumes the
existence of quasiparticle  exitations for the spherically
symmetric core.

The goal of this paper is to develop a self-consistent method
allowing to calculate effects of core polarization in a field of
valent electron within the framework of secondary quantized
electron-"hole"\ formalizm.


\section{Wave function of many-electron system
}
Let an arbitrary function
$\psi$ to depend on coordinates 
$\vec r_i$, $i,\ldots, n$ of
$n$ electrons: 
$\psi = \psi(\vec r_1, \vec r_2,\ldots, \vec
r_n)$. To be of a wave function for a system from  $n$ electrons the function
$\psi(\vec r_{\alpha_1}, \vec
r_{\alpha_2},\ldots, \vec r_{\alpha_n})$
has to satisfy the Pauli principle, i.~e. to be antisymmetric over
space coordinates.
Here the index $\alpha_i,\ i=1,\ldots,n$ runs 
$\{1,2,\ldots, n\}$ so that 
$\alpha_i \neq\alpha_j$ for 
$i\neq j$. It can be achieved if one represents this function as
\begin{eqnarray}
\psi(\vec r_{\alpha_1}, \vec r_{\alpha_2},\ldots, \vec r_{\alpha_n})=
\epsilon (P_{\alpha_1 \alpha_2 \ldots \alpha_n})
\psi(\vec r_1, \vec r_2,\ldots, \vec r_n),
\label{fermion-func}
\end{eqnarray}
where $P_{\alpha_1 \alpha_2 \ldots \alpha_n}$ is a permutation:
\begin{eqnarray}
P_{\alpha_1 \alpha_2 \ldots \alpha_n}=\left(
\begin{array}{cccc}
1     &           2  &  \ldots & n\\
\alpha_1 & \alpha_2 & \ldots & \alpha_n
\end{array}
\right)
\end{eqnarray}
which converts $1$ into 
$\alpha_1$, $2$ into 
$\alpha_2$ and so on. 
A symbol $\epsilon(P)$ denotes a number equal to 
$+1$ if  the permutation is even, and a number equal to
$-1$ if  the permutation is odd. 

It is easy to prove that the functions (\ref{fermion-func})
satisfy the following equality:
\begin{eqnarray}
\sum_{\{\alpha_i\}_{i=1}^n}
\psi(\vec r_{\alpha_1}, \vec r_{\alpha_2},\ldots, \vec r_{\alpha_n})=
0.
\label{fermion-func-cycle}
\end{eqnarray}
To do it we divide up the left side of the expression
(\ref{fermion-func-cycle}) into functions 
so that one  describes the spin configuration as
$\{ \uparrow\uparrow \ldots \uparrow \uparrow|
\downarrow \downarrow \downarrow \ldots \downarrow \}
$, and another one describes the spin configurations:
$\{ \uparrow \uparrow \ldots \uparrow \downarrow|
\uparrow \downarrow \downarrow \ldots \downarrow \}
$,
$\{ \uparrow\uparrow \ldots \uparrow \downarrow|
\downarrow \uparrow \downarrow \ldots \downarrow \}
$,
$\{ \uparrow\uparrow \ldots \uparrow \downarrow|
\downarrow \downarrow \uparrow \ldots \downarrow \}
$, $\ldots$,
$\{ \uparrow\uparrow \ldots \uparrow \downarrow|
\downarrow \downarrow \downarrow \ldots \downarrow \uparrow\}
$, with the sign minus
$'' - ''$.
Graphically this partitioning is shown in fig.~1 and is written
mathematically as
\begin{eqnarray}
\psi (\vec r_1,  \ldots ,\vec r_{k-1}, \vec r_k|
\vec r_{k+1},   \vec r_{k+2}\ldots , \vec r_n)=
\psi (\vec r_1,  \ldots ,\vec r_{k-1}, \vec r_{k+1}|
\vec r_,k   \vec r_{k+2}\ldots , \vec r_n)+\ldots
\nonumber \\
+
\psi (\vec r_1,  \ldots ,\vec r_{k-1}, \vec r_{k+l}|
\vec r_{k+1}, \ldots , \vec r_{k+l-1},\vec r_k ,\vec r_{k+l+1}, \ldots,\vec r_n)
+\ldots
\nonumber \\
+
\psi (\vec r_1,  \ldots ,\vec r_{k-1}, \vec r_n|
\vec r_{k+1}, \ldots , \vec r_{k-n},\vec r_k). \label{cycle-simmenry}
\end{eqnarray}

$$\{ \underbrace{ \uparrow\uparrow \uparrow \uparrow}|
\underbrace{\downarrow \downarrow \downarrow \downarrow\downarrow }\}=
\{ \underbrace{\uparrow \uparrow \uparrow \downarrow}|
\underbrace{\uparrow \downarrow \downarrow \downarrow \downarrow }\}
+
\{\underbrace{ \uparrow\uparrow  \uparrow \downarrow}|
\underbrace{\downarrow \uparrow \downarrow \downarrow \downarrow} \}
+
 \{ \underbrace{\uparrow\uparrow \uparrow \downarrow}|
\underbrace{\downarrow \downarrow \uparrow \downarrow \downarrow }\}
+\ldots +
\{ \underbrace{\uparrow\uparrow  \uparrow \downarrow}|
\underbrace{\downarrow \downarrow \downarrow \downarrow \downarrow \uparrow}\}
$$

\vspace{-4mm}\hspace{0.6cm}{\small $k$ \hspace{0.5cm}$n-k$
\hspace{1.1cm} $k$ \hspace{0.5cm}$n-k$ \hspace{0.9cm} $k$
\hspace{0.5cm}$n-k$ \hspace{1.1cm} $k$ \hspace{0.5cm}$n-k$
\hspace{1.9cm} $k$ \hspace{0.5cm}$n-k$}

{\small 
Fig. 1. Graphics represents  the property of cyclic symmetry for
a wave function of electron.
}

\vspace{4mm}



But the wave function situated in the right side of symbolical
expression in fig.~1, describes a configuration obtained by a cyclic
permutation of electrons from the configuration at the left in
fig.~1. Thus we have proved, that the electron function  is symmetric
with respect to the cyclic permutation and a mathematical notation
of this property of cyclic symmetry is the expression
(\ref{cycle-simmenry}).

The set of functions (\ref{fermion-func}) is a basic set to
construct the wave function of a many-electron system. Slater
determinants possess  the properties such as functions from the
introduced basic set.

Further we use a method of secondary quantization to establish a
Hartree - Fock equation describing a single-electron state in the
self-consistent field, within the framework of electron-hole
formalism.

\section{Electron-hole formalism
}

Hamiltonian $H$ for the many-electron system reads:
\begin{eqnarray}
H(\vec r_1,\ldots , \vec r_n)   = \sum_{i=1}^n \left[
-{\hbar ^2\over 2m} \nabla ^2(\vec r_i) +V(\vec r_i)\right] +
\sum_{i> j=1}^n {e^2\over |\vec r_i -\vec r_j|}
= \sum_{i=1}^n H(\vec r_i)  +
\sum_{i> j=1}^n V(\vec r_i -\vec r_j)
  , \label{Schrodinger-Hamiltonian1}
\end{eqnarray}
where $\nabla ^2$ is the Laplacian, 
$V(\vec r_i)$ is the potential energy 
of $i$-th 
electron in a field of atomic nucleus or nuclei, $ V(\vec r_i -\vec r_j)$
is the Coulomb interaction potential. 

Let us go into representation of secondary quantization where a single-particle
state is given by creation operators
${\widehat \psi}^\dag (  x_i)$
and annihilation operators ${\widehat \psi} (  x_i)$ of  $i$-th
Fermi particle with generalized coordinates
$x_i=\{\vec r_i, t_i,\sigma_i\}$ 
being its radius-vector $\vec r_i$, 
time $t_i$ and spin 
$\sigma_i$. 
These operators satisfy the commutation relations \cite{SecondquantFock}:
\begin{eqnarray}
{\widehat \psi} (  x') {\widehat \psi}^\dag (x)+
 {\widehat \psi}^\dag (x) {\widehat \psi} (  x')=\delta (x-x')
 \label{permutation1} \\
{\widehat \psi} (  x') {\widehat \psi}(x)+
 {\widehat \psi} (x) {\widehat \psi} (  x')=0.  \label{permutation2}
\end{eqnarray}
Now one can introduce an operator of "hole"\ creation in the following way.
Since the wave function of the system can be written in the form
\begin{eqnarray}
{\widehat \psi}^\dag _{(n-k)\downarrow}
(  \vec r_n, \ldots ,  \vec r_{k+2}, \vec r_{k+1})
{\widehat \psi}^\dag _{k\uparrow}
(\vec r_k ,\vec r_{k-1},  \ldots ,\vec r_1)
|0\rangle \nonumber \\
= [\widehat \psi _{n\downarrow}(\vec r_n)+
\widehat \psi _{(n+1)\uparrow}(\vec r_{n+1})]
{\widehat \psi}^\dag _{(n-k+1)\downarrow}
(  \vec r_{n+1}, \ldots ,  \vec r_{k+2}, \vec r_{k+1})
{\widehat \psi}^\dag _{k\uparrow}
(\vec r_k ,\vec r_{k-1},  \ldots ,\vec r_1)
|0\rangle
\nonumber \\
=[\widehat \psi _{n\downarrow}(\vec r_n)+
\widehat \psi _{(n+1)\uparrow}(\vec r_{n+1})]
|\psi _1,\ldots, \psi _{n+1} \rangle , \label{hole}
\end{eqnarray}
the operator
$[\widehat \psi _{n\downarrow}(\vec r_n)+
\widehat \psi _{(n+1)\uparrow}(\vec r_{n+1})]$ is
a creation operator of a "hole"\ by $n$-th electron. 
Here ${\widehat \psi}^\dag _{n\downarrow}
(  \vec r_n \ldots ,  \vec r_{k+2}, \vec r_{k+1})
{\widehat \psi}^\dag _{k\uparrow}(\vec r_k ,\vec r_{k-1},  \ldots ,\vec r_1)$
is a secondary quantized wave function of the system describing
the configuration from $k$ electrons with spin "up"\
and $n-k$ electrons with spin "down"\ ,
$n=2k+1$, i.e. 
there is one unpaired electron;
$|0\rangle$ is a vacuum state. 
The wave function (\ref{hole}) describes systems with outer unpaired electron
or with a "hole", having spin "up"\
$\uparrow$ by outer electron. 

Let us assume that  core polarization happens at the  moment
$t $. Then in this moment  $t $ the secondary quantized wave
function of system can be obtained as a result of cyclic
permutation $P ^{(cycl)} (t) $:
\begin{eqnarray}
P^{(cycl)}(t)
[\widehat \psi _{n\downarrow}(\vec r_n)+
\widehat \psi _{(n+1)\uparrow}(\vec r_{n+1})]
|\psi _1,\ldots, \psi _{n+1}\rangle . \label{core-polarization}
\end{eqnarray}
Since the operators $P^{(cycl)}(t)$ and 
$[\widehat \psi _{n\downarrow}(\vec r_n)
+ \widehat \psi _{(n+1)\uparrow}(\vec r_{n+1})]$ commute with each other
then according to the definition of permutation operator which is shown
in fig.~1 the expression
(\ref{core-polarization}) can be rewritten as 
\begin{eqnarray}
P^{(cycl)}(t)
[\widehat \psi _{n\downarrow}(\vec r_n)+
\widehat \psi _{(n+1)\uparrow}(\vec r_{n+1})]
|\psi _1,\ldots, \psi _{n+1}\rangle \nonumber \\
=
[\widehat \psi _{n\downarrow}(\vec r_n)+
\widehat \psi _{(n+1)\uparrow}(\vec r_{n+1})]
P^{(cycl)}(t)
|\psi _1,\ldots, \psi _{n+1}\rangle  \nonumber \\
=
\left[\sum_{m=1}^k c_{nm}(t)\widehat \psi _{m\downarrow} (\vec r_m,t) +
\sum_{m=k+1}^n c_{nm}(t)\widehat \psi _{(m+1)\uparrow}(\vec r_{m+1},t)\right]
P_m^{(cycl)} |\psi _1,\ldots, \psi _{n+1}\rangle
\nonumber \\
= \sum_{m=1}^n c_{nm}(t)\widehat \psi  (x_m)
P_m^{(cycl)} |\psi _1,\ldots, \psi _{n+1}\rangle
\label{core-polarization},
\end{eqnarray}
where $m$-th term of the sum describes 
a "hole"\   which is created at $m$-th core electron after the cyclic permutation
$P_m^{(cycl)}$ in the sense of the expression
(\ref{cycle-simmenry}),
a matrix $|c_{nm}(t)|$ converts the creation operator of the "hole"\ by
$n$-th electron into the creation operator of the "hole"\ by
$m$-th electron. 

According to  the property of cyclic symmetry
(\ref{cycle-simmenry}) one can formally write the expression
\begin{eqnarray}
[\widehat \psi _{n\downarrow}(\vec r_n)+
\widehat \psi _{(n+1)\uparrow}(\vec r_{n+1})]
=
\sum_{m=1}^n c_{nm}(t)\widehat \psi (x_m)P_m^{(cycl)}
 \label{core-polarization1}.
\end{eqnarray}
It follows from here that the sum
$\sum_{m=1}^n c_{nm}(t)\widehat \psi(x_m)$ has to satisfy the same quantum
equations of motion as for
$[\widehat \psi _{n\downarrow}(\vec r_n)+
\widehat \psi _{(n+1)\uparrow}(\vec r_{n+1})]$.
Heisenberg equation of motion for the "hole"\ creation operator
$[\widehat \psi _{n\downarrow}(\vec r_n)+
\widehat \psi _{(n+1)\uparrow}(\vec r_{n+1})]$ reads:
\begin{eqnarray}
{\hbar \over \imath }{\partial \over \partial t}
[\widehat \psi _{n\downarrow}(\vec r_n)+
\widehat \psi _{(n+1)\uparrow}(\vec r_{n+1})]=
\left[
[\widehat \psi _{n\downarrow}(\vec r_n)+
\widehat \psi _{(n+1)\uparrow}(\vec r_{n+1})],
\widehat{\mathcal{H}}
\right];                 \label{Geisenberg-motion-eq}\\
\widehat{\mathcal{H}}=\left(
\sum_{i=1}^n \int  \widehat  H(x_i)d \vec r_i +
\sum_{i> j=1}^n \int \int \widehat  V(x_i,x_j)
d \vec r_i d \vec r_j
\right)
\\
\widehat H(x_i)={\widehat \psi}^\dag (x_i) H(\vec r_i) {\widehat \psi }(x_i);\\
\widehat  V(x_i,x_j)= {1\over 2}
{\widehat \psi}^\dag (x_j) {\widehat \psi}^\dag (x_i)
V(\vec r_i -\vec r_j) {\widehat \psi}(x_i) {\widehat \psi} (x_j)
\end{eqnarray}
where $[\widehat A, \widehat B]$ is a commutator of
operators $\widehat A$ and 
$\widehat B$, $\widehat{\mathcal{H}}$ is the Hamiltonian operator in the
secondary quantization formalism
which is obtained by a secondary quantization procedure \cite{SecondquantFock}
from the operator (\ref{Schrodinger-Hamiltonian1}).

Substituting the expression
(\ref{core-polarization1}) into  the equation of motion 
(\ref{Geisenberg-motion-eq}) one gets
\begin{eqnarray}
{\hbar \over \imath }{\partial \over \partial t}
\sum_{m=1}^n c_{nm}(t)\widehat \psi (x_m)P_m^{(cycl)}=\sum_{m,i=1}^n c_{nm}(t)
\Bigg\{ \int d \vec r_i
\nonumber\\
\times
\widehat \psi (x_m)
 \left[
{\widehat \psi}^\dag (x_i) H(\vec r_i) {\widehat \psi }(x_i)+
\sum_{j=1}^n \int
{ d \vec r_j\over 2}
{\widehat \psi}^\dag (x_j) {\widehat \psi}^\dag (x_i)
V(\vec r_i -\vec r_j) {\widehat \psi}(x_i) {\widehat \psi} (x_j)
\right]
\nonumber\\
-
 \left[
{\widehat \psi}^\dag (x_i) H(\vec r_i) {\widehat \psi }(x_i)+
\sum_{j=1}^n \int
{ d \vec r_j\over 2}
{\widehat \psi}^\dag (x_j) {\widehat \psi}^\dag (x_i)
V(\vec r_i -\vec r_j) {\widehat \psi}(x_i) {\widehat \psi} (x_j)
\right]\widehat \psi (x_m)
\Bigg\}P_m^{(cycl)}                   \label{Geisenberg-motion-eq1}
\\
\nonumber \mbox{for 
}j<i
\end{eqnarray}
for the "hole"\ creation operator
$\sum_{m=1}^n c_{nm}(t)\widehat \psi (x_m)$ which depends on configuration.
Using the commutation rules (\ref{permutation1}, \ref{permutation2}) for
quantized fermion fields one transforms eq.~
(\ref{Geisenberg-motion-eq1}) to the form:
\begin{eqnarray}
{\hbar \over \imath }{\partial \over \partial t}
\sum_{m=1}^n c_{nm}(t)\widehat \psi (x_m)P_m^{(cycl)} =\sum_{m,i=1}^n c_{nm}(t)
\Bigg\{ \int d \vec r_i
\nonumber\\
\times
\widehat \psi (x_m)
 \left[
\delta (x_i-x_m) H(\vec r_i) {\widehat \psi }(x_i)+
\sum_{j=1}^n \int
{ d \vec r_j\over 2}
\delta (x_j-x_m) {\widehat \psi}^\dag (x_i)
V(\vec r_i -\vec r_j) {\widehat \psi}(x_i) {\widehat \psi} (x_j)
\right]
\nonumber\\
-
 \left[
\sum_{j=1}^n \int
{ d \vec r_j\over 2} \delta (x_i-x_m)
{\widehat \psi}^\dag (x_j)
V(\vec r_i -\vec r_j) {\widehat \psi}(x_i) {\widehat \psi} (x_j)
\right]\widehat \psi (x_m)
\Bigg\}P_m^{(cycl)}                 \label{Geisenberg-motion-eq2}
\\
\nonumber \mbox{for 
}j<i
\end{eqnarray}
where $\delta (x_k-x_m)$ is the Dirac 
$\delta$-function. 
Differentiating over time at the left side and integrating over the Dirac
$\delta$-function 
at the right side of eq.~(\ref{Geisenberg-motion-eq2}) one gets finally
\begin{eqnarray}
\left({\hbar \over \imath }
{\partial \ln c_{nm}(t) \over \partial t}-
\widehat \epsilon\ \widehat { \mbox{I}} \right) \widehat \psi (x_m)P_m^{(cycl)}=
\Bigg( H(\vec r_m) {\widehat \psi }(x_m)+{1\over 2}
\sum_{i=1}^n \int { d \vec r_i}
\nonumber\\
\times \left(
 {\widehat \psi}^\dag (x_i)
V(\vec r_i -\vec r_m) {\widehat \psi}(x_i) {\widehat \psi} (x_m)
-
{\widehat \psi}^\dag (x_i)
V(\vec r_m -\vec r_i) {\widehat \psi}(x_m) {\widehat \psi} (x_i)\right)
\Bigg) P_m^{(cycl)}
\nonumber\\
=\Bigg( H(\vec r_m) {\widehat \psi }(x_m)-
\sum_{i=1}^n \int { d \vec r_i}
 {\widehat \psi}^\dag (x_i)
V(\vec r_i -\vec r_m) {\widehat \psi}(x_m) ({\widehat \psi} (x_m)\delta_{mi})
\Bigg) P_m^{(cycl)}
,                 \label{Geisenberg-motion-eq3}
\end{eqnarray}
where $\widehat { \mbox{I}}$ is the unity operator,
$\widehat \epsilon$  is the operator of "hole"\ energy
as
$$
{\widehat \psi }(x_m)={\widehat \psi }(\vec r_m,\sigma_m)
\mbox{exp}(-\imath\widehat \epsilon\ \widehat { \mbox{I}} t/\hbar);
$$
and the right side of eq.~(\ref{Geisenberg-motion-eq3}) is rewritten
taking into account matrix multiplication rules.

Now we can
find the equations, describing single-particle state neglecting
correlations in movement of electrons relative to each  other. To
examine the configuration shown in fig.~1, we shall assume that
electrons with spins "up" \ move independent on electrons with
spins "down".
 In other words their movement are not correlated.
Therefore the wave function of such configuration is factorized in the
following way:
\begin{eqnarray}
{\widehat \psi}^\dag _{(n-k)\downarrow}
(  \vec r_n, \ldots ,  \vec r_{k+2}, \vec r_{k+1})
{\widehat \psi}^\dag _{k\uparrow}
(\vec r_k ,\vec r_{k-1},  \ldots ,\vec r_1)
|0\rangle
=
|\psi _1,\ldots, \psi _{n}\rangle \nonumber\\
=
|  \psi_n, \ldots ,  \psi_{k+2}, \psi_{k+1}\rangle
|\psi_k ,\psi_{k-1},  \ldots ,\psi_1)
\rangle
\nonumber\\
={\widehat \psi}^\dag _{(n-k)\downarrow}
(  \vec r_n, \ldots ,  \vec r_{k+2}, \vec r_{k+1})|0\downarrow\rangle
{\widehat \psi}^\dag _{k\uparrow}
(\vec r_k ,\vec r_{k-1},  \ldots ,\vec r_1)
|0\uparrow\rangle
.\label{decoupling}
\end{eqnarray}
From here it follows an expansion for the vacuum state $|0\rangle$
\begin{eqnarray}
|0\rangle
=
|0\downarrow\rangle |0\uparrow\rangle \equiv
|0,\sigma_m \rangle |0,-\sigma_i\rangle
,\label{decoupling1}
\end{eqnarray}
which means that
$|0\rangle$ consists of non-occupation states
with spins "down"\ $|0\downarrow\rangle$ and  with spins "up"\
$|0\uparrow\rangle$.

Hermitian conjugation of eq.~(\ref{Geisenberg-motion-eq3}) has the form:
\begin{eqnarray}
\left(\imath\hbar
{\partial \ln c_{nm}(t) \over \partial t}-
\widehat \epsilon^\dag\ \widehat { \mbox{I}} \right)
P_m^{(cycl)}\widehat \psi^\dag (x_m)\nonumber\\
=
P_m^{(cycl)}\Bigg( H(\vec r_m) {\widehat \psi }^\dag(x_m)-
\sum_{i=1}^n \int { d \vec r_i}
{\widehat \psi}^\dag (x_m)
V(\vec r_i -\vec r_m) ({\widehat \psi}^\dag (x_m)\delta_{mi}){\widehat \psi}(x_i)
\Bigg)\nonumber\\
=
P_m^{(cycl)}\Bigg( H(\vec r_m) {\widehat \psi }^\dag(x_m)-
\sum_{i=1}^n \int { d \vec r_i}
{\widehat \psi}^\dag (x_m)
V(\vec r_i -\vec r_m) {\widehat \psi}^\dag (x_i){\widehat \psi}(x_i)
\Bigg)
.                 \label{Geisenberg-motion-eq4}
\end{eqnarray}
Acting with the hermitian conjugate equation
(\ref{Geisenberg-motion-eq4}) on
the found non-occupied states $|0\sigma_m\rangle $ in
the vacuum state 
(\ref{decoupling1}),
one gets the following equation
\begin{eqnarray}
\left(\imath\hbar
{\partial \ln c_{nm}(t) \over \partial t}-
\widehat \epsilon^\dag\ \widehat { \mbox{I}} \right)
P_m^{(cycl)}\widehat \psi^\dag_{\sigma_m} (\vec r_m)
|0\sigma_m\rangle |0,-\sigma_i\rangle \nonumber\\
=
P_m^{(cycl)}\Bigg( H(\vec r_m) {\widehat \psi }^\dag_{\sigma_m}(\vec r_m)-
\sum_{i=1}^n \int { d \vec r_i}
{\widehat \psi}^\dag _{\sigma_m}(\vec r_m)
{\widehat \psi}^\dag_{\sigma_i}  (\vec r_i)
V(\vec r_i -\vec r_m) {\widehat \psi}_{-\sigma_i}(\vec r_i)
\Bigg)|0\sigma_m\rangle|0,-\sigma_i\rangle
.            \nonumber\\    \label{Geisenberg-motion-eq5}
\end{eqnarray}
According to the expression (\ref{cycle-simmenry})
the operation of permutation 
$P_m^{(cycl)}$ entered into eq.~
(\ref{Geisenberg-motion-eq5}) is written in the explicit form as
\begin{eqnarray}
P_m^{(cycl)} {\widehat \psi}^\dag _{\sigma_m}(\vec r_m)
{\widehat \psi}^\dag_{\sigma_i} (\vec r_i)
{\widehat \psi}_{-\sigma_i}(\vec r_i)
={\widehat \psi}^\dag _{\sigma_i}(\vec r_m)
{\widehat \psi}^\dag_{-\sigma_i} (\vec r_i)
{\widehat \psi}_{\sigma_m}(\vec r_i)-
{\widehat \psi}^\dag _{\sigma_m}(\vec r_m)
{\widehat \psi}^\dag_{-\sigma_i} (\vec r_i)
{\widehat \psi}_{\sigma_i}(\vec r_i)
.\label{cycl-symmetry-three-term}
\end{eqnarray}
Substitution of the explicit expression for
$P_m^{(cycl)}$ (\ref{cycl-symmetry-three-term}) into
eq.~(\ref{Geisenberg-motion-eq5}) gives the following
equation: 
\begin{eqnarray}
\left(\imath\hbar
{\partial \ln c_{nm}(t) \over \partial t}-
\widehat \epsilon^\dag\ \widehat { \mbox{I}} \right)
\widehat \psi^\dag_{\sigma_m} (\vec r_m)
|0\sigma_m\rangle |0,-\sigma_i\rangle 
=
\Bigg( H(\vec r_m) {\widehat \psi }^\dag_{\sigma_m}(\vec r_m)-
\sum_{i=1}^n \int { d \vec r_i}
 \nonumber\\
\times (
{\widehat \psi}^\dag _{\sigma_i}(\vec r_m)V(\vec r_i -\vec r_m)
{\widehat \psi}^\dag_{-\sigma_i} (\vec r_i)
{\widehat \psi}_{\sigma_m}(\vec r_i)-
{\widehat \psi}^\dag _{\sigma_m}(\vec r_m)V(\vec r_i -\vec r_m)
{\widehat \psi}^\dag_{-\sigma_i} (\vec r_i)
{\widehat \psi}_{\sigma_i}(\vec r_i))
\Bigg)\nonumber\\
\times
|0\sigma_m\rangle|0,-\sigma_i\rangle
.            \nonumber\\    \label{Geisenberg-motion-eq6}
\end{eqnarray}
Multiplying on the left side of eq.~(\ref{Geisenberg-motion-eq6}) 
by the vector $\langle 0,-\sigma_i|$
one gets 
\begin{eqnarray}
 \imath\hbar
{\partial \ln c_{nm}(t)\over \partial t}
\widehat \psi^\dag_{\sigma_m}(\vec r_m) |0\sigma_m\rangle
- \langle 0,-\sigma_i|
\widehat \epsilon^\dag\ \widehat { \mbox{I}}|0,-\sigma_i\rangle
\widehat \psi^\dag_{\sigma_m} (\vec r_m)
|0\sigma_m\rangle  
=
 H(\vec r_m) {\widehat \psi }_{\sigma_m}(\vec r_m)|0\sigma_m\rangle
\nonumber\\
  -
\sum_{i=1}^n \int { d \vec r_i}
{\widehat \psi}^\dag _{\sigma_i}(\vec r_m)|0\sigma_m\rangle
V(\vec r_i -\vec r_m)
\langle 0,-\sigma_i|{\widehat \psi}^\dag_{-\sigma_i} (\vec r_i)
{\widehat \psi}_{\sigma_m}(\vec r_i)|0,-\sigma_i\rangle
 \nonumber\\
 + \sum_{i=1}^n \int { d \vec r_i}
{\widehat \psi}^\dag _{\sigma_m}(\vec r_m) |0\sigma_m\rangle
V(\vec r_i -\vec r_m)\langle 0,-\sigma_i|
{\widehat \psi}^\dag_{-\sigma_i} (\vec r_i)
{\widehat \psi}_{\sigma_i}(\vec r_i)
|0,-\sigma_i\rangle
,           \nonumber\\
 \label{Geisenberg-motion-eq7}
\end{eqnarray}
as $\langle 0,-\sigma_i|0,-\sigma_i\rangle = 1$.
If one introduces the following designation:
${\widehat \psi}^\dag_{\sigma_j}(\vec r_k)|0\sigma_j\rangle
\equiv {\psi}_{j}(x_k)$ and represents the unity operator
in the explicit form:
$$\widehat { \mbox{I}}=\sum_{j=1}^n
{\widehat \psi}^\dag_{\sigma_j}(\vec r_k)|0\sigma_j\rangle
\langle 0\sigma_j |{\widehat \psi}_{\sigma_j}(\vec r_k)\equiv
\sum_{j=1}^n P_j
,
$$
then eq.~(\ref{Geisenberg-motion-eq7}) can be rewritten as
\begin{eqnarray}
 \imath\hbar
{\partial \ln c_{nm}(t) \over \partial t} \psi_{m} (x_m)-
\sum_{j=1}^n\widehat \epsilon^\dag  P_j
 \psi_{m} (x_m)  
=
 H(\vec r_m) \psi _{m} (x_m)
 \nonumber\\
 + \sum_{i=1}^n \int { d \vec r_i}\left(
{ \psi} _{m}(x_m) V(\vec r_i -\vec r_m)
{ \psi}^*_{i} (x_i)
{ \psi}_{i}(x_i)-
{\psi} _{i}(x_m)
V(\vec r_i -\vec r_m){ \psi}^*_{i} (x_i)
{ \psi}_{m}(x_i)\right)
.          \label{Geisenberg-motion-eq8}
\end{eqnarray}
Eq.~(\ref{Geisenberg-motion-eq8}) 
taking at initial time $t=0$
is the equation describing single-particle state $ {\psi}_{m}(x_m)$:
\begin{eqnarray}
\left[ H(\vec r_m)
 +  \hat V^{sc} (x_m) - \hat \Sigma ^x(x_m) \right] \psi_m (x_m)
 =\left(\epsilon_m(0)+ \sum_{j=1}^n\widehat \epsilon^\dag  P_j
 \right) \psi_m (x_m),
 \label{hartry-fock-eqs}
\end{eqnarray}
where the differentiation over time $t$ taking at initial time
$t=0$ is designated as
$\epsilon_m(0)$:
$$
\epsilon_m(0) = - \left.\imath\hbar
{\partial \ln c_{nm}(t) \over \partial t}\right|_{t=0},
$$
$\hat V^{sc}$ and 
$ \hat \Sigma ^x $ are the Coulomb and exchange interactions,
respectively:
\begin{eqnarray}
\hat V^{sc}(x_i)\psi_n (x_i)=\sum_{m=1}^n
\int \psi_m^* (x_j) v(|\vec r_i - \vec r_j|)\psi_m (x_j)
\ d   \vec r_j \psi_n (x_i)
,
\label{coulon}\\
\hat \Sigma ^x(x_i) \psi_n (x_i)
=\sum_{m=1}^n
\int \psi_m^* (x_j) v(|\vec r_i - \vec r_j|)\psi_n (x_j)
\ d  \vec r_j \psi_m (x_i). 
\label{exchange}
\end{eqnarray}
Physical meaning of the operators (\ref{coulon},
\ref{exchange}) becomes evident if one rewrites them in terms of
spinless electronic density
$\rho(\vec r, \vec r')$ and supposes that the interaction
$v$ is the Coulomb one:
\begin{eqnarray}
&\rho(\vec r, \vec r')={1\over 2}\sum_{m=1}^{n-1}\left(
 \psi_m^* ( \vec r ,\sigma )
\psi_m ( \vec r' , -\sigma )+
\psi_m^* (  \vec r ,-\sigma)
\psi_m ( \vec r', \sigma )
\right)
\nonumber\\
&=\sum_{m=1}^{(n-1)/2}
\psi_m^* (  \vec r)
\psi_m (  \vec r'),\ v =e^2/|\vec r - \vec r'|.
\end{eqnarray}
From here  it follows that the operator
$\hat V^{sc}$ represents electrostatic interaction of one electron with
the electron density produced by remaining
$n-1$ electrons
and  electrostatic self-action (s.a.):
\begin{eqnarray}
\hat V^{sc}(x_i)\psi_n (x_i) 
=
\sum_\sigma \sum_{m=1}^{(n-1)/2}
\int \psi_m^* ( \vec r_j) v(|\vec r_i - \vec r_j|)
\psi_m (  \vec r_j)
\ d   \vec r_j \psi_n (  \vec r_i)
+\mbox{s.a.}
\nonumber \\
=2 \int \ d   \vec r_j
{e^2 \rho(\vec r_j, \vec r_j)\over |\vec r_i - \vec r_j|}
\psi_n (x_i)
+\mbox{s.a.}
\end{eqnarray}
Analogously one obtains that the operator $ \hat \Sigma ^x $ gives
quantum exchange with exchange self-action
(s.a.):
\begin{eqnarray}
\hat \Sigma ^x(x_i) \psi_n (x_i)
=\sum_{m=1}^{n-1}
\int d  r_j \psi_m^* (  \vec r_j, \sigma_j)
v(|\vec r_i - \vec r_j|)\psi_n (  \vec r_j, \sigma_j)
 \psi_m (  \vec r_i, \sigma_i) +\mbox{s.a.} \nonumber \\
 ={1\over 2}\sum_{m=1}^{n-1}\int \int d  r_j d \sigma_j
 \left(
 \psi_m^* (  \vec r_j ,\sigma_j )
\psi_m ( \vec r_i , -\sigma _j) \delta(\sigma_j - \sigma_i )\right.\nonumber \\
\left. +
\psi_m^* (  \vec r_i ,-\sigma_j)
\psi_m ( \vec r_j, \sigma_j)\delta(\sigma_j - \sigma_i )
\right)
v(|\vec r_i - \vec r_j|)\psi_n ( \vec r_j, \sigma_j)+\mbox{s.a.}
\nonumber \\
 =\int \ d   \vec r_j
{e^2 \rho(\vec r _j, \vec r _i)\over |\vec r_i - \vec r _j|}
\psi_n (x_j)
+\mbox{s.a.}
\label{exchange1}
\end{eqnarray}
Since the operators $\hat V^{sc}$ and 
$ \hat \Sigma ^x $ into the expression 
(\ref{hartry-fock-eqs}) are subtracted from each other
then the self-acting terms (s.a.) entering into it
are mutually cancelled. 

Let us suppose that there exists a representation in which the energy operator
of a "hole"\
$\widehat\epsilon^\dag$ entering into
eq.~(\ref{hartry-fock-eqs})
is diagonalized $\widehat\epsilon ^\dag= \epsilon (k_i)I$,
where 
$I$ is an unitary matrix.
The replacing $\vec r_m\to \vec r_i$ and the taking into account of
the diagonalization condition into 
eq.~(\ref{hartry-fock-eqs}) allow 
to describe the core polarization as a quasiparticle exitation
with the energy $\epsilon (k_i)$
which stationary state is determined as
\begin{eqnarray}
\left[ H(\vec r_i)
 +  \hat V^{sc} (k_ix_i) - \hat \Sigma ^x(k_ix_i) \right] \psi_m (k_ix_i)
 =\left(\epsilon_m(0)+ \sum_{j=1}^n\widehat \epsilon^\dag  P_j
 \right) \psi_m (k_ix_i),
 \label{hartry-fock-eqs1}
 \\
 \widehat\epsilon ^\dag= \epsilon (k_i)I.
\end{eqnarray}

{\it Approximation of a "frozen"\ atom
}

Let us call by approximation of a "frozen"\ atom the
calculations with assumption that almost all time a "hole"\ is near
$m$-th electron which is located in a point with radius-vector
$\vec r_i$.
Since the  energy operator $\hat\epsilon of $"hole"\
by $m$-th electron in the point 
$\vec r_i$ can be of a work only which has to be perfomed to shift an electron
occupying
the "hole"\ near $m$-th electron to
$j$-th 
non-occupied orbital of atom "frozen"\ in the moment, then
$\hat\epsilon $ is equal to: 
\begin{eqnarray}
\hat\epsilon  ^{(m)}_{ji} = -( \epsilon_{j} - \epsilon_{i}^{(m)}).
\label{hole-energy}
\end{eqnarray}
Substituting the expression (\ref{hole-energy}) into 
eq.~(\ref{hartry-fock-eqs1}) one gets
\begin{eqnarray}
\left[ H(\vec r_i)
 +  \hat V^{sc} (x_i) - \hat \Sigma ^x(x_i) \right] \psi_m (x_i)
 =\left(\epsilon_m(0)- \sum_{j=1}^n( \epsilon_{i}^{(m)} -\epsilon_{j} ) P_j
 \right) \psi_m (x_i).
 \label{hartry-fock-eqs2}
\end{eqnarray}

The approximation
of valent electron $ \psi_v $ or  the approximation of a "frozen" \
atomic core appears if there exists an outer valent $v $-th
electron located in a point with a radius - vector $ \vec r_v $
such that it always is outside limits of atomic core: $ \vec r_v>
\vec R_c $, where $ | \vec R_c | $ is a radius of the "frozen" \
core. Existence of outer electron means that one of electrons can
not be placed inside of the core. It is possible if the core  has no
"holes". Since in this approximation the core has no "holes"\ then
eq.~(\ref {hartry-fock-eqs2}) can be rewritten as
\begin{eqnarray}
\left[ H(\vec r_i)
 +  \hat V^{sc} (x_i) - \hat \Sigma ^x(x_i) \right] \psi_m (x_i)
 =\left(\epsilon_m(0)- (1- \delta_{ic})
 \sum_{j=1}^n( \epsilon_{i} -\epsilon_{j} ) P_j
 \right) \psi_m (x_i), 
 \label{hartry-fock-eqs3}
\end{eqnarray}
the valent orbital is the orbital determined by the expression
\begin{eqnarray}
\psi_v = (1-\sum_c P_c)\psi_m(x_i), \quad i\neq c .
\end{eqnarray}
Here index "$c$"\ labels core electrons. 
It is easy see that eq.~(\ref{hartry-fock-eqs3}) is  the equation
for Phillips - Kleinman pseudo-potential $V^{PK}$
\cite{Phillips}, \cite{Weeks}, \cite{Dixon} in the case of wave function
depending on both coordinates and spin variables of electron.
Therefore eq.~(\ref{hartry-fock-eqs1}) and 
eq.~(\ref{hartry-fock-eqs2}) are  generalization of
eq.~(\ref{hartry-fock-eqs3}) and with their help one can describe many-electron
systems beyond the framework where the method of pseudo-potential is valid.

\section{
Secondary quantized  density matrix}
We rewrite (\ref{hartry-fock-eqs1})
in the representation of the Dirac ket (bra) - vectors:
\begin{eqnarray}
\hat h(k) | n; k \rangle + \sum_{m=1, m\neq n}^N
\int \delta(k-k') \ dk' (| n; k \rangle \langle  m; k' | v(kk')| m; k'\rangle
\nonumber \\
+ (| n; k' \rangle \delta_{nm})\langle  m; k' | v(kk')
(| m; k\rangle \delta_{mn} ))
\nonumber \\
-
\sum_{m=1}^N
\int | m; k \rangle \langle  m; k' | v(kk')| n; k'\rangle  \delta(k-k') \ dk'
= | n; k \rangle (\epsilon_n(0)+ \epsilon_n(k))
\label{moment-Hartry-Fock-eq}
\end{eqnarray}
where
$k_i\equiv \{\vec{k}_i, \sigma_i\}$, $\hat h(k)$ is
a momentum representation of the non-perturbed hamiltonian,

\noindent
$v(kk')=\int d \vec r d\vec r'
|\vec r\rangle \langle  \vec k\cdot \vec r |
v(|\vec r - \vec r'|)
| \vec k'\cdot \vec r '\rangle \langle \vec r' |$
is a momentum representation of the Coulomb interaction operator,
$\delta(k-k')$
is the Dirac 
$\delta $-function manifesting the presence of the law of conservation of
momentum.

Let us introduce projective operators $\hat\rho ^{mn}_{kk'}$
\cite{Grush-Var-Princip}:
\begin{eqnarray}
\hat \rho ^{mn}_{kk'} \equiv | m; k'\rangle \langle  n; k|.
\label{secondary-quant-projector}
\end{eqnarray}
and
express 
eq.~(\ref{moment-Hartry-Fock-eq}) through these operators 
$\hat \rho ^{mn}_{kk'}$.
%
For this purpose, Eq.~(\ref {moment-Hartry-Fock-eq}) is multiplied
  from the right by bra-vector $ \langle n; k | $. Then, additional summating over
$n $ and integrating over $dk $  lead to the  equation:
\begin{eqnarray}
\sum_{n=1}^N \int dk\ \hat h(k) | n; k \rangle \langle  n; k|+ \sum_{n=1}^N
\int \int \delta(k-k')\ dk\ dk'
\times \nonumber \\
\times
| n; k \rangle \sum_{m=1}^N \langle  m; k' | v(k,k')| m; k'\rangle
\langle  n; k|
+ \int \int \delta(k-k')\ dk\ dk' \times \nonumber \\
\times
\left(\sum_{m=1}^N| n; k '\rangle \delta_{nm}\langle  m; k' |\right) v(kk')
\left(\sum_{ n=1}^N| m; k\rangle \delta_{mn}\langle  n; k| \right)
-\nonumber \\
-
\int \int \sum_{m=1}^N | m; k \rangle \langle  m; k' | v(kk')
\sum_{n=1}^N| n; k'\rangle \langle  n; k|
\delta(k'-k) d(-k) d(-k')
\nonumber \\
= \int dk \sum_{n=1}^N \langle  n; k|n; k \rangle (\epsilon_n(0)
+ \epsilon_n(k)).
\nonumber \\
\label{moment-Hartry-Fock-eq1}
\end{eqnarray}
We see that the first and second terms in the left side of the equation
(\ref {moment-Hartry-Fock-eq1}) are traces of a matrix
representation of operators $ \hat \rho \hat h $ and $ \hat \rho
^* \hat \rho v $, and the third and fourth terms in the left side of
equation (\ref {moment-Hartry-Fock-eq1}) are mutually canceled.
Hence, it means that using  normalization of the function $
|n; k \rangle $: $ \int dk\langle n; k|n; k \rangle =1 $, we
obtain the following equation:
\begin{eqnarray}
\mbox{Sp}   \hat \rho \hat h
+ \mbox{Sp} \hat \rho ^* \hat \rho v
=\epsilon_n(0)N +\int dk \sum_{n=1}^N \langle  n; k|n; k \rangle
\epsilon_n(k).
\label{moment-Hartry-Fock-eq2}
\end{eqnarray}
Using properties of the projective operators $\hat\rho ^{mn}_{kk'}$:
$\left(\hat\rho ^{mn}_{kk'}\right)^*=\hat\rho ^{mn}_{kk'}$ and 
$\left(\hat\rho ^{mn}_{kk'}\right)^2=\hat\rho ^{mn}_{kk'}$,
one can transform 
eq.~(\ref{moment-Hartry-Fock-eq2}) to the form: 
\begin{eqnarray}
\mbox{Sp}   \hat \rho (\hat h +  v)
=\epsilon_n(0)N +\int dk \sum_{n=1}^N \langle  n; k|n; k \rangle
 \epsilon_n(k)= \epsilon_n(0)N + \epsilon.
\label{moment-Hartry-Fock-eq3}
\end{eqnarray}

Let us elucidate  physical sense of introduced projective
operators $ \hat\rho ^ {mn} _ {kk '} $.  The
energy of a quasiparticle $ \epsilon $ is  in the right side of
Eq.~(\ref {moment-Hartry-Fock-eq3}) up to the constant $
\epsilon_n {(0)} N $. From here it follows  that the operator $
\hat\rho ^ {mn} _ {kk '} $ allows to calculate  the energy $
\epsilon $ of quasiparticle excitations. It means that the
expression (\ref {moment-Hartry-Fock-eq3}) is  nothing else but
a procedure of average over density matrix. Since the averaging with
the help of the operator $ \hat\rho ^ {mn} _ {kk '} $ yields the
energy $ \epsilon $ of quasiparticle, this operator is a secondary
quantized density matrix.

Thus, it is proved that eq.~(\ref{hartry-fock-eqs1}) can be considered
as the equation describing the state of quasiparticle and
determining its energy with accuracy up to the constant $\epsilon_n {(0)}N$.
Since for description of single-electron state it is necessary to take into
account the presence of "hole", then the quasiparticle state describes
the electron-hole pair.

\section{Green function for single-particle state
}
It follows also from Eq.~(\ref {moment-Hartry-Fock-eq3})  that the
quantity $ \epsilon_n {(k_i)} $ can be interpreted as an
eigenvalue of the Hamiltonian for a quasiparticle excitation
without taking into account  interaction of quasiparticles.
Therefore, the equation (\ref {moment-Hartry-Fock-eq3}), written
in the formalism of the density matrix $ \hat\rho ^ {(0)} _ {nn '; kk
'} \equiv \hat\rho ^ {mn} _ {kk '} $ can be rewritten in the
formalism of wave functions in coordinate representation and in the
limit of large $N $, $N\to \infty $ in the following way:
\begin{eqnarray}
 \left(\imath {\partial \over \partial t }-(\hat h +  \Sigma^{x}+V^{sc} )\right)
 \sum_n   \hat\rho ^{(0)}_{nn'; rr'}
=\lim_{N\to\infty}  (-\epsilon_n(0))N \delta_{rr'},
\label{moment-Hartry-Fock-eq4}
\end{eqnarray}
where property $ \int dk\langle  n; kr|n; kr' \rangle =\delta_{rr'}$
has been used;
$\delta_{rr'}$ is a delta symbol.
Since the energy $ \epsilon_n (0) $ of  the bound one-electron
state is negative: $ \epsilon_n (0) <0 $, the right side of Eq.~
(\ref {moment-Hartry-Fock-eq4}) represents   itself the Dirac
$\delta$-function $ \delta(r-r')$. It allows to write Eq.~(\ref
{moment-Hartry-Fock-eq4}) as:
\begin{eqnarray}
 \left(\imath {\partial \over \partial t }-\hat h^{HF} \right)
 \sum_n   \hat\rho ^{(0)}_{nn'; r,r'}
=\delta(r-r'),
\label{moment-Hartry-Fock-eq5}
\end{eqnarray}
where  
$\hat h^{HF}=(\hat h +  \overline\Sigma^{x}+V^{sc})$,
$\overline \Sigma^{x}=-\Sigma^{x}$.
Eq.~(\ref{moment-Hartry-Fock-eq5}) is the equation for the Green
function. It means that in the secondary quantized representation the operator
\begin{eqnarray}
\hat G_1^{(0)}(n'; r,r') = \sum_n \hat\rho ^{(0)}_{nn'; r,r'}
\end{eqnarray}
possesses properties of non-perturbed Green function.

So, the quasiparticle excitation determined by the Hamiltonian $
\hat h ^ {HF} $ can be considered as a free particle whose
equation of motion is the equation (\ref{moment-Hartry-Fock-eq5}).

In many-body problem, in particular, in calculations of
energy-band of crystal structure   a contribution given by the interaction of
an electromagnetic field with matter plays  an essential role. To take into
account many-particle effects due to
correlated motion of an electron we should describe the system
by self-consistent solutions of the non-stationary equation
\begin{eqnarray}
 \imath {\partial\Psi (t) \over \partial t }= \hat H \Psi (t)
\label{non-stationary-quant-motion-eq}
\end{eqnarray}
%
where $\hat H$ is the Schr\"odinger hamiltonian in a
non-relativistic case or a Dirac hamiltonian in  relativistic one.

We have proved that for the secondary quantized representation the
operator $ \rho $ looks like $ \hat \rho = | \hat \psi \rangle
\langle \hat\psi | $ and possesses the properties of the Green
function $G_1 $. Therefore the sum $ \hat G_1 $ over $n $ from
elements of the matrix $ \hat\rho ^ {nn '} _ {kk '} $ for the
secondary quantized density matrix $ \hat \rho $ describing an
interacting particle satisfies the Dyson equation in
nonrelativistic case or to the  Schwinger - Dyson equation in
relativistic case:
\begin{eqnarray}
G_1(1;2)= G_1^{(0)}(1;2)+\int d3\ d4\ G_1^{(0)}(1;3) \hat \Sigma (3,4) G_1(4;2)
\label{Shwinger-Dayson-eq}
\end{eqnarray}
where $G_1^{(0)}(1;2)$ is the free Green function,
$\hat \Sigma (3,4)$ is the self-energy operator:
$\hat \Sigma = {\overline \Sigma^{x}}+\hat \Sigma^{c}$,
$\hat \Sigma^{c}$ is  correlation interactions,
representing itself a part of the self-energy which describes the
many-particle effects.
Here numerical labels for the arguments are used: $ \{r_1, t_1 \}
= x_1\equiv 1 $, etc. Acting on Eq.~(\ref {Shwinger-Dayson-eq}) by
the operator $ \imath {\partial \over \partial t}-\hat h ^ {HF} $
and using the equation of motion for the free particle
(\ref{moment-Hartry-Fock-eq5}) we get the equation for the perturbed
Green function as
\begin{eqnarray}
\left[ \imath {\partial \over \partial t }-\hat h^{HF}(r_1)\right]G_1(n';1,2)-
\int d3 \hat \Sigma^{c} (n';1,3) G_1(n';3,2) 
=
(-\epsilon_n(0))N \delta_{r_1r_2}.
\label{Perturbed-Green-func-eq}
\end{eqnarray}
Rewriting Eq.~(\ref{Perturbed-Green-func-eq}) in the formalism of
wave functions one gets
\begin{eqnarray}
\left[\imath {\partial \over \partial t }-\hat h^{HF}(r_1)\right]\psi_{n}
(k_1r_1) -
\int d \vec r_2 \hat \Sigma^{c} (n;1,2) \psi_{n}(k_1r_2)=
(-\epsilon_n(0)) \psi_{n}
(k_1r_1).
\label{Perturbed-wave-func-eq1}
\end{eqnarray}
Since the expression:  
$\imath {\partial \psi_{n}\over \partial t} =\epsilon_n(k_1)$
takes place, then
Eq.~(\ref{Perturbed-wave-func-eq1}) yields  
the Hartree - Fock equation taking into account of interacting
quasiparticles
\begin{eqnarray}
 \hat h^{HF}(r_1)\psi_{n}
(k_1r_1) + \int d \vec r_2 \hat \Sigma^c (n;1,2) \psi_{n}(k_1r_2)
=(\epsilon_n(0)+\epsilon_n(k_1)) \psi_{n}
(k_1r_1).
\label{Perturbed-Hartry-Fock-eqs}
\end{eqnarray}

Let us define a mass operator $\widehat {\Delta M} $ as: 
\begin{eqnarray}
\widehat {\Delta M}\psi_n (k_1 r_1)
 = \int d \vec r_2 \hat \Sigma^c (n;1,2) \psi_{n}(k_1r_2).
\end{eqnarray}
Within the framework of the concept of quasiparticle excitations it
is possible to represent the operator $ \widehat {\Delta M} $ in
the diagonal form:
\begin{eqnarray}
\widehat {\Delta M}\psi_n (k_i r_i)
 =-(\Delta M_n(0)+ \Delta M_n(k_i)) \psi_n (k_i r_i),
\end{eqnarray}
the eigenvalue of mass operator possesses the
property:

\noindent
$\Delta M_n(k_i)=\Delta M_n(-k_i)$. Here 
$\Delta M_n(0)$ is an eigenvalue of mass operator 
$\widehat {\Delta M}$ in the limit 
$\vec k \to 0$.

From here it follows   physical sense of $ \widehat {\Delta M} $.
It determines an effective mass of the quasiparticle and an
efficient bottom of the energy band:
\begin{eqnarray}
 \hat h^{HF}(r_1)\psi_{n}(k_1r_1) =
(\tilde \epsilon_n(0)+\tilde \epsilon_n(k_1))\psi_{n}(k_1r_1)\equiv \nonumber \\
 \equiv
\left[(\epsilon_n(0)+\Delta M_n(0))+(\epsilon_n(k_1)+\Delta M_n(k_1))\right]
 \psi_{n}(k_1r_1).
\label{Perturbed-Hartry-Fock-eqs1}
\end{eqnarray}
Extremum of energy band  $\mbox{Extr} E_n(k_i)$
in the presence of interaction between quasiparticles is determined by
the following expression:
\begin{eqnarray}
\mbox{Extr} E_n(k_i)  =  (\Delta M_n(0)+\epsilon_n(0))N.
\label{variation-princip3}
\end{eqnarray}

Let us consider the Green function normalized per unit volume $V=1 $
so that   average energy in $V $ is equal to the energy of
single-particle state and $N=1 $. If $-\epsilon_n (0) \to \infty $
then Eq.~(\ref {Perturbed-Green-func-eq}) describes  propagation
of single particle and should be rewritten as
\begin{eqnarray}
\left[ \imath {\partial \over \partial t }-\hat h^{HF}(r_1)\right]G_1(n';1,2)-
\int d3 \hat \Sigma^c (n';1,3) G_1(n';3,2)= \nonumber \\
=
(-\epsilon_n(0)) \delta_{r_1r_2}.
\label{Particle-Green-func-eq}
\end{eqnarray}
Then according to the definition of the Green
functions we have the following expression for the energy $
\epsilon_n (0) $:
\begin{eqnarray}
 -\epsilon_n(0)=C - a_n, \quad C \to \infty; \label{Particle-energy}
\end{eqnarray}
where $a_n$ is a finite quantity.
Hence, since the energy is counted off from an arbitrary value,
Eq.~ (\ref {variation-princip3}) yields the following expression for the
reference points $
\epsilon (0) ^ {\pm}_n $ of  quasiparticle energy and
antiquasiparticle energy
\begin{eqnarray}
  \epsilon(0)^{\pm}_n\equiv \pm a_n
  =  \left( \mbox{Extr} \tilde{E}(k_1)\mp \Delta {M}_n(0) \right )/2.
\label{zone-reference}
\end{eqnarray}
Here one took into account that
$N=1$; 
an extremum of zone is redefined as
$\mbox{Extr}\tilde{E}(k_1)= \mbox{Extr}{E}_n(k_1)-C$,
the sign $\{\pm\}$ in the left-hand side denotes  cases of quasiparticles and
antiquasi\-par\-tic\-les, respectively;
and the energy of particles in the pair is counted off from zero level.
From the expression (\ref{zone-reference}) one gets
that $a_n$ is the energy which is required to create a pair from
quasiparticle  and antiquasiparticle
when $k_1=0$ because
\begin{eqnarray}
a_n=(\epsilon(0)^{+}_n - \epsilon(0)^{-}_n)/2.
\label{gap}
\end{eqnarray}

Because of an additional term $ \tilde \epsilon_n (0) $ in the
right-hand side of  Eq.~(\ref{Perturbed-Hartry-Fock-eqs1}) we,
generally speaking, cannot examine the left-hand side as a
Hamiltonian operator of  quasiparticles system acting on a
corresponding wave function and as a consequence,  can not
construct a basis set of single-particle states of the problem. However,
further we show, that $ \hat h^{HF} $ is a Hamiltonian of an
electron - hole pair.

{\it Non-relativistic case 
}

In non-relativistic limit one can examine  quantum systems which
are characterized by a small value of $\Delta { M}_n(0)$:
\begin{equation}
\Delta {M}_n (0)\to 0. \label{light-electron}
\end{equation}
It means that weak many-particle effects occur and, accordingly,
we can speak about a "light" \  electron. The equality (\ref
{zone-reference})  occurs under condition of
(\ref{light-electron}) only in the case if $a_n=0 $. From here  it
follows, that the energy $a_n $ of a pair is equal to zero. In other
words,  the energy is not expended to create an electron - hole
pair.

Substituting Eqs.~(\ref{Particle-energy}), (\ref{light-electron})
into
(\ref{Perturbed-Hartry-Fock-eqs1}) 
and taking into account the
condition
$a_n=0$ 
one gets the Schr\"odinger equation in the form
\begin{eqnarray}
 \hat h^{HF}(r_1)\psi_{n}(r_1) =
\tilde {\tilde \epsilon}_n \psi_{n}(r_1),
\label{Perturbed-Hartry-Fock-eqs2}
\end{eqnarray}
which describes the quasiparticle - antiquasiparticle pair (a
non-relativistic electron - hole pair). Here $\tilde {\tilde
\epsilon}_n={\tilde \epsilon}_n -C$.

Since the  energy $a_n $, expended on  creation of a pair, equals
to zero we have proved that the variable $ \tilde {\tilde
\epsilon} _n $ can be understood as the energy of an electron -
hole pair. Therefore, Eq.~(\ref {Perturbed-Hartry-Fock-eqs2}) has
a group of dynamic symmetry, which algebra is \textsf
{so(3)~x~so(3)} $ \sim $\textsf{so(4)} if to neglect  exchange
interaction. As is known, a nonrelativistic hydrogen-like atom
possesses such a symmetry. Hence, we have proved that to calculate
quasiparticle states in  non-relativistic case it is possible
to use a basis set of states of a nonrelativistic hydrogen-like
atom.

However, for a heavy electron $ \Delta M_n (0) \ge 1 $ according
to the formula (\ref{gap}) we always have
\begin{equation}
a_n=-\Delta {M}_n (0)/2 \label{heavy-electron}
\end{equation}
and, hence, there is no an   equation such as the Schr\"odinger
one for its description. From here we conclude that the heavy
electron can not be examined in nonrelativistic limit.

{\it Relativistic case 
}

Let us generalize  the proposed approach to relativistic case. To
do it we substitute Eqs.~(\ref{Particle-energy}) and
(\ref{heavy-electron}) into
(\ref{Perturbed-Hartry-Fock-eqs1}) and
let $ n$  tends to $ n\to \infty $:
\begin{eqnarray}
 \hat h^{HF}(r_1)\psi_{n}(k_1r_1) =
\left({\Delta { M}_n (0)\over 2}+\tilde {\tilde \epsilon}_n(k_1)
\right)\psi_{n}(k_1r_1),
\quad n\to \infty.
\label{Quasirelativistic-eq}
\end{eqnarray}
Then, one can assume that the operator ${\partial \over \partial
t}- \hat h^{HF}$ in Eq.~(\ref{Quasirelativistic-eq}) is a quasirelativistic
hamiltonian  written in the implicit form in the Hartee - Fock approximation.

From  consideration carried out above it  follows that the desired
relativistic equation of motion should describe a charged
composite system from a pair of particles and have the dynamic
symmetry \textsf{SO(4)}. A spin of a given quantum system should be
equal 1 as motion of a hole is a motion of   an electron in
many-particle positively charged matrix.
In \cite{Gurs&Grush-DocBGUIR2003} the equation of motion of a relativistic
charged vector boson has been found and it was shown, that it describes a
relativistic hydrogen-like atom.
The relativistic charged vector-boson appears as a composite system
with a corresponding spectrum of masses and in quasirelativistic
limit $n\to \infty $ its energy $E_1 $ is determined by the
expression:
\begin{eqnarray}
 E_1\approx {m\over 2} -{m\gamma ^2 \over 2{n }^2} - {m \gamma ^4\over 8 n ^3}
 { \left({4\over   |k|  }-{3\over   n  }\right)}-
 {m \gamma ^6\over 8 n ^4}
 { \left({3\over   n^2  }-{8\over  n  |k|  }+{4\over  k^2  }\right)}
 +O(\gamma^8).\nonumber \\
\ \label{boson-energy}
\end{eqnarray}
Here $k$ is a quantum number of relativistic angular momentum:
$k=-l,\ l+1$, $l$ is a quantum number of orbital moment.
Comparison of right-hand sides of formulas (\ref{Quasirelativistic-eq}) and
(\ref {boson-energy}) yields that $
\Delta M _ {\infty} \equiv \lim _ {n\to \infty} \Delta M_n (0) =m
$ is a rest mass  $m $ of an electron. Hence,
Eq.~(\ref{Quasirelativistic-eq}) is an equation of motion for a
relativistic  electron - hole pair with a reduced mass $\Delta M _
{\infty}/2=m/2 $ which, apparently, is a relativistic charged
vector-boson considered in quasirelativistic limit $n\to \infty $.

\section{Conclusion 
}


The method of Hartree - Fock  self-consistent field in
non-relativistic case or of Dirac - Fock self-consistent field  in
relativistic case is applicable for  description of valent
electron and polarization of atomic core only within the framework
of the perturbation theory with the pseudopotential method as a
zero-order approximation. To describe effects of  polarization of
electron density distribution in the case of intensive electromagnetic
interactions with a atomic core it is necessary to take into
account correlation interaction which is neglected in the above mentioned
 method.
The method offered in this paper allows to describe effects of
polarization not only for a atomic core in a field of valent
electron, but also polarization of atom as the whole in the
electron-hole formalism.


\end{document}